\documentclass[fleqn,usenatbib]{mnras}

\usepackage{newtxtext,newtxmath}

\usepackage[T1]{fontenc}

\DeclareRobustCommand{\VAN}[3]{#2}
\let\VANthebibliography\thebibliography
\def\thebibliography{\DeclareRobustCommand{\VAN}[3]{##3}\VANthebibliography}

\usepackage{graphicx}	
\usepackage{amsmath}	

\title[NS-WD PNe]{The evolutionary route to form planetary nebulae with central neutron star – white dwarf binary systems}

\author[I. Ablimit \& N. Soker]{
Iminhaji Ablimit,$^{1,2}$\thanks{E-mail: iminhaji@nao.cas.cn (IA)}
and Noam Soker$^{3}$\thanks{soker@physics.technion.ac.il (NS)}
\\
$^{1}$CAS Key Laboratory for Optical Astronomy, National Astronomical Observatories, Chinese Academy of Sciences, Beijing 100012, China\\
$^{2}$Department of Astronomy, Kyoto University, Kitashirakawa-Oiwake-cho, Sakyo-ku, Kyoto 606-8502, Japan\\
$^{3}$Department of Physics, Technion - Israel Institute of Technology, Haifa 32000, Israel
}

\pubyear{2023}

\begin{document}
\label{firstpage}
\pagerange{\pageref{firstpage}--\pageref{lastpage}}
\maketitle

\begin{abstract}
We present a possible evolutionary pathway to form planetary nebulae (PNe) with close neutron star (NS)-white dwarf (WD) binary central stars. By employing the binary population synthesis technique we find that the evolution involves two common envelope evolution (CEE) phases and a core collapse supernova explosion between them that forms the NS. Later the lower mass star engulfs the NS as it becomes a red giant, a process that leads to the second CEE phase and to the ejection of the envelope. This leaves a hot horizontal branch star that evolves to become a helium WD and an expanding nebula. Both the WD and the NS power the nebula. The NS in addition might power a pulsar wind nebula inside the expanding PN.  From our simulations we find that the Galactic formation rate of NS-WD PNe is $1.8 \times 10^{-5}\,{\rm yr}^{-1}$ while the Galactic formation rate of all PNe is $0.42 \,{\rm yr}^{-1}$. There is a possibility that one of the observed Galactic PNe might be a NS-WD PN, and a few NS-WD PNe might exist in the Galaxy. The central binary systems might be sources for future gravitational wave detectors like LISA, and possibly of electromagnetic telescopes.    
\end{abstract}

\begin{keywords}
(stars:) binaries (including multiple): close -- stars: evolution -- (stars:) white dwarfs -- (ISM:) planetary nebulae: general -- (stars:) supernovae: general - stars: late-type
\end{keywords}



\section{Introduction}
\label{sec:Introduction}

A  planetary nebula (PN) is a late evolutionary phase of low and intermediate-mass stars, i.e., zero age main sequence (ZAMS) mass in the range of $\simeq 0.8 - 8.0 M_\odot$.
The basic structure of a PN is of a hot, effective temperature of $T_{\rm eff} \ga 3\times 10^4 {\rm K}$, central star that is the remnant of an asymptotic giant branch (AGB) stellar progenitor or of a red giant (RG) star, and an expanding nebula that was the envelope of the AGB or RG progenitor of the PN (e.g., \citealt{Kwok1983, TweedyKwitter1994, Soker2006, Schonberneretal2007, Coxetal2012, GuerreroDe Marco2013, KwitterHenry2022}). Most PNe come from AGB stars, with only a small fraction from RG stars (e.g., \citealt{Hillwigetal2017, Jonesetal2020, Jonesetal2022, Jonesetal2023}). The central star evolves to become a white dwarf (WD). 

Thousands of PNe with various morphologies (i.e. elliptical, round, bipolar or butterfly, lacking any symmetry and termed messy PNe), sizes, ionization properties, central star properties and chemical abundances have been discovered in the Milky Way (e.g., \citealt{Greig1971, Manchadoetal2000, Stanghellinietal2002, Corradietal2003, Drewetal2005, Parkeretal2005, Miszalskietal2008, Sahai2011, Sabinetal2014, Kronbergeretal2014, Parkeretal2016}). This large variety of PN properties stimulated many observational and theoretical studies (e.g., \citealt{Morris1987, Icke1989, TweedyKwitter1994, SokerRappaport2000, Miszalskietal2009, FrewParker2010, DeMarcoetal2015, Hillwigetal2016, JonesBoffin2017, Jacobyetal2021}, to list a small fraction out of hundreds of papers). 

Most studies attribute the variety of PN morphologies to the interaction of the AGB or the RG progenitors with a companion, stellar or sub-stellar, including mass transfer, launching of jets, and common envelope evolution (e.g. \citealt{Morris1981, Morris1987, Soker1997, SokerRappaport2001, DeMarco2009, GarciaSegura2018, Frank2018, Ondratschek2022, GarciaSegura2022}).
About 20\% of central stars in known PNe are close binaries (e.g., \citealt{Bond2000, Miszalskietal2009}), with increasing number of central binary stars detection by deeper sky surveys (e.g., \citealt{Bakeretal2018}), like by the Kepler space telescope and by Gaia survey (e.g., \citealt{BoffinJones2019, Jacobyetal2021, Chornay2021, ChornayWalton2022}). 
Mass-accreting WD systems are reported as the central stars in some known PNe (\citealt{Bodeetal1987, Guerreroetal2004, Wessonetal2008, Kahabkaetal2008, Munarietal2013, MaitraHaberl2022}). We here use the term WD to indicate also central stars that are not yet WDs but evolving to become WDs. Some WDs might accrete mass from their non-degenerate companions (\citealt{Hamannetal2003, Guerreroetal2019, Jonesetal2019}), a process that might explain  some puzzles, like the luminosity function of PNe (\citealt{Ciardullo2016, Davisetal2018, Souropanis2023}). Merger of a WD companion with the core might even set a type Ia supernova explosion during the PN phase
(e.g., \citealt{TsebrenkoSkoer2013, TsebrenkoSkoer2015, Cikotaetal2017, Chiotellisetal2020, Chiotellisetal2021}).

In close compact star (i.e., WD or neutron star)--non-degenerate star binaries, the mass loss during the stable Roche-lobe overflow (RLOF) mass transfer and common envelop evolution (CEE) caused by the unstable RLOF mass transfer may form PNe with a rich variety of properties. To our knowledge, the physical processes of the evolution of compact star binaries that form PNe and the possible outcomes are not thoroughly explored yet.
In this work we employ the binary population synthesis (BPS) method to study binary evolution, including RLOF, mass transfer, mass ejection, supernova mechanism with kick, and CEE, to explore the formation of a specific type of peculiar PNe, i.e., PNe that have central NS-WD binary systems. 
In section \ref{sec:BinaryEvolution} we introduce the main physical ingredients of the \textsc{bse} BPS code. In section \ref{sec:Results} we present the typical evolutionary pathway to form PNe with central binary systems composed of a newborn WD in a close orbit with a NS; we term these NS-WD PNe. We also determine the formation rate of NS-WD PNe and the properties of their central binary system, which might power pulsar wind nebula and in the far future might be a gravitational wave source. We conclude in section \ref{sec:Summary}.

\section{Binary evolution setup}
\label{sec:BinaryEvolution}

We use the BPS code \textsc{bse} (\citealt{Hurleyetal2002, KielHurley2006}; see also \citealt{Ablimitetal2016, AblimitMaeda2018} and  \citealt{Ablimitetal2022})  to simulate the evolution of a large binary population ($10^7$ systems) starting as ZAMS stars.
The \textsc{bse} code is an extension of the single stellar evolution code \textsc{sse} which is based on \cite{Hurleyetal2000}. We adopt the updated version of  \textsc{bse} and \textsc{sse} to study the formation of very rare and peculiar NS-WD PNe.
We here list the significant changes and processes in the updated code (\citealt{Ablimitetal2016, AblimitMaeda2018, Ablimit2021} and  \citealt{Ablimitetal2022}). 
 
For the initial primary (the initially more massive star) masses  
we adopt the distributions of \cite{Kroupaetal1993} (see also \citealt{Kroupa2001}) 
\begin{equation}
f(M_1) = \left\{ \begin{array}{ll}
0 & \quad  \textrm{${M_1/M_\odot} < 0.1$}\\
0.29056{(M_1)}^{-1.3} & \quad \textrm{$0.1\leq {M_1/M_\odot} < 0.5$}\\
0.1557{(M_1)}^{-2.2} & \quad  \textrm{$0.5\leq {M_1/M_\odot} < 1.0$}\\
0.1557{(M_1)}^{-2.7} & \quad  \textrm{$1.0\leq {M_1/M_\odot} \leq 100$}.
\end{array} \right.
\end{equation}
A flat/uniform distribution is used for the mass ratio of
the binary $\rm q = M_2/M_1$ (see also \citealt{Ferrario2012, MoeDiStefano2017}) to derive the initial mass of the secondary star $M_2$.
The initial orbital separations (semi-major axes) are simulated to be flat in logarithm scale (e.g., \citealt{opik1924, MoeDiStefano2017}).
We take a thermal distribution of the eccentricity $e$ ($f(e)=2e$; \citealt{Heggie1975}) in the range of $0 \le e <1$. 

For hot stars (O and B stars in different stages) the stellar wind mass loss rate prescription in the code is
revised with the wind model of \cite{Vinketal2001}.
 The luminous blue-variable wind is calculated as
$\dot{M}_{\rm lbv} = 10^{-4} f_{\rm lbv} M_\odot\,{\rm yr}^{-1}$ with $f_{\rm lbv} = 1.5$. 
For stripped Helium stars, Wolf-Rayet stars and other types of stars, model 2 in \cite{AblimitMaeda2018} and the wind mass-loss prescriptions of \cite{VinkdeKoter2005} and of \cite{Hurleyetal2000} are adopted.

CCSN explosion mechanisms and natal kicks are crucial but are not well-known to be accurately incorporated in stellar and binary evolution. Two different supernova prescriptions for determining the remnant mass are introduced in \cite{Fryeretal2012}, neutrino-driven and convection-enhanced. In calculating the final remnant mass they take into account material that is falling back onto the compact object formed in the CCSN explosion itself. 
The rapid remnant-mass model of \cite{Fryeretal2012} allows for explosions in a short timescale and produces the 
remnant distributions with the mass gap between NSs and black holes (BHs). On the other hand, the delayed supernova model has the explosions in a relatively longer timescale and does not reproduce the mass gap between NSs and BHs. We adopt the rapid remnant-mass model of \cite{Fryeretal2012} to determine the remnant masses (see \citealt{Mandeletal2021} for a different prescription). The orbital change (the binary can even be disrupted) due to the ejection of mass during the CCSN explosion and due to the NS natal kick are also included.  
We draw the SN kick (imparted on the newborn NS) velocities in cases of iron-core collapse CCSN from a Maxwellian distribution with a dispersion parameter of $\sigma = 265 \rm{km\,s^{-1}}$ (\citealt{Hobbsetal2005}).
The dispersion parameter for NSs born in electron-capture SNe is $\sigma = 40 \rm{km\,s^{-1}}$ (\citealt{Dessartetal2006}). The mass range for electron-capture supernovae is adopted from \cite{Podsiadlowskietal2004}. The NS mass range\footnote{The NS mass could reflect the SN explosion mechanisms (e.g., \citealt{Pejchaetal2012})} is between $\simeq 1$ and $\simeq 2.0 M_\odot$.

The CEE (e.g., \citealt{Paczynski1971}) is a key phase in the evolution of binary systems to form compact objects as gravitational wave sources, interesting 
transients like SNe and peculiar astronomical objects (e.g., \citealt{PortegiesYun1998, Hurleyetal2002, Belczynskietal2008, Izzardetal2012, Toonenetal2012, deminkBel2015, wangetal2015, Kruckowetal2016, GiacobboMapelli2018, Andrewsetal2018, Eldridgeetal2018, Mapelli2019, Ablimit2021, Olejaketal2021, Broekgaardenetal2021, Marchantetal2021, Hamersetal2021, Kruckowetal2021, Ablimitetal2022, vansonetal2022, MandelBroekgaarden2022, Koroletal2022, Tanikawaetal2022, Rileyetal2022, Tranietal2022, Fragosetal2023, GagnierPejcha2023, Ohetal2023, vanetal2023}, and see \cite{ropkede2023} for the very recent review).
The mass ratio has a decisive role in determining whether a CEE occurs. If the mass ratio (i.e. donor/accretor) is larger than a critical mass ratio,  $M_{\rm donor}/M_{\rm accretor}>q_{\rm c}$, then mass transfer is dynamically unstable and a CEE takes place. 
If the donor star is on the MS or crosses the Hertzsprung gap we use $q_{\rm c} = q_{\rm const} = 4.0$, while if the donor is on the first giant branch (i.e. RG) or on the AGB we use
\begin{equation}
q_{\rm c} = 0.362 + \frac{1}{3(1 - {M_{\rm c}}/M)},
\end{equation}
where $M$ and $M_{\rm c}$ are the total stellar mass and core mass of the donor star (the giant), respectively. 
For donors (primaries) that are He stars we take $q_{\rm c} = 3.0$ for helium-rich MS stars and $q_{\rm c} = 0.784$ for helium-rich giants (see \citealt{Hurleyetal2002} for more details).
The mass-transfer efficiency determines how much of the transferred gas is accreted by the accretor and how much escapes from the binary. We consider the gas to escape (or ejected from) the binary system when it is no longer affected by the binary. We assume that material that is ejected from the system carries the specific angular momentum of the accretor (e.g., \citealt{Hurleyetal2002}). 

The \textsc{bse} BPS code uses the standard energy conservation prescription (the alpha-CEE prescription; e.g, \citealt{Paczynski1971}) 
\begin{equation}
E_{\rm{bind}} = {\alpha_{\rm CE}} \Delta E_{\rm orb} \ ,
\label{eq:EbindEenv}
\end{equation}
where $E_{\rm{bind}}$ and $\Delta E_{\rm orb}$ are the binding energy of the envelope and the change in the orbital energy during the CEE phase, respectively. Equation (\ref{eq:EbindEenv}) defines the CEE parameter $\alpha_{\rm CE}$. In the specific model of \cite{Webbink1984} that we
adopt here the binding energy of the envelope is parameterized as
\begin{equation}
E_{\rm{bind}} = - \frac{GM_1 M_{\rm{en}}}{\lambda {R}_1},
\end{equation}
where $M_1$, $M_{\rm{en}}$ and ${R}_1$ are the total mass, envelope mass and radius of the giant star, respectively.  We use constant values for the CEE efficiency and for the binding energy parameters as $\alpha_{\rm CE} = 1.0$ and $\lambda = \lambda_{\rm w}$ as the one in \cite{Ablimit2021} (see also \citealt{AblimitMaeda2018}), respectively. Whether the two stars merge completely or survive the CEE to continue their evolution critically depends on these two parameters. 

We set solar metallicity (0.02) as the initial metallicity of the two stars. 
Other physical parameters of the initial MS-MS binaries are the same as in the default prescriptions in \cite{Hurleyetal2000, Hurleyetal2002}. 
The different prescriptions that different BPS codes use (that include many simplifications) and the uncertainties in the codes affect the results (see \citealt{Ablimitetal2022} for different models and related outcomes). We must bear in mind these large uncertainties when discussing the results. We here mainly present the new formation pathway for a very rare type of objects, namely, PNe with a NS-WD close binary system in their center that possibly powers a pulsar wind nebula. Note that the NS becomes a reborn pulsar as it accretes mass from the envelope of the giant star. The accretion through an accretion disk amplifies magnetic fields and spins-up the NS. 

\section{Results}
\label{sec:Results}

\begin{figure*}
\centering
\includegraphics[trim=0cm 0cm 0.0cm 0.0cm ,clip, scale=0.8]{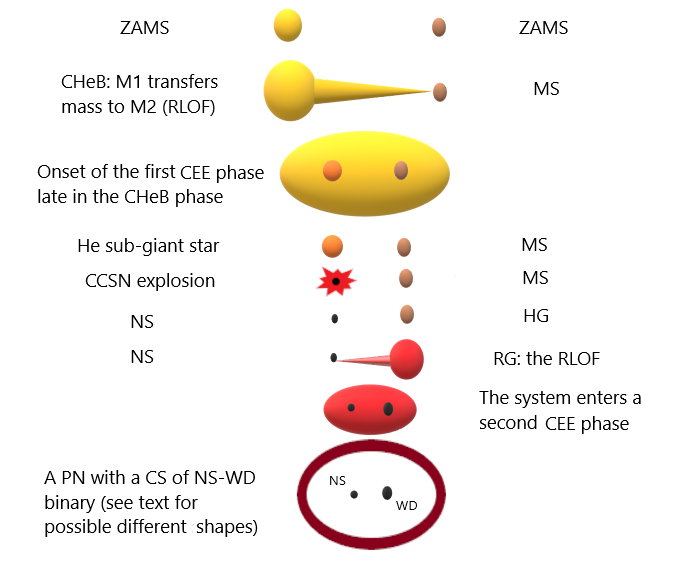}
\caption{A schematic description of a typical evolutionary pathway to form a NS-WD PN, i.e., a PN with a central NS-WD binary system at its center. In all cases the NS-WD that we find are post-RG PNe. The drawn elliptically-shaped nebula is a schematic one, as the nebula might be bipolar and contain a pulsar wind nebula in the inner region. Abbreviations: CCSN: core collapse supernova; CEE: common envelope evolution; CHeB: core helium burning; HG: Hertzsprung gap (star); MS: main sequence (star); NS: neutron star; RG: red giant; RLOF: Roche lobe overflow; WD: white dwarf; ZAMS: zero-age main sequence.}
\label{Fig:schematic} 
\end{figure*}

In Figure \ref{Fig:schematic} we schematically present the typical evolutionary pathway for the formation of PNe with a NS-He WD binary at their center. The primary star in the initial MS-MS binary system in a relatively wide orbit is sufficiently massive to end in a CCSN explosion that leaves a NS remnant. Along its evolution the primary star overflows its Roche lobe when it is in the core helium burning (CHeB) phase. This drives an unstable mass transfer process and the binary system finds itself in the first CEE phase. The MS secondary star might merge with the core of the primary giant star or survive the CEE. We are interested in the second possibility.  
At the end of the first CEE phase the primary star becomes a stripped-envelope helium (He) sub-giant star and the secondary star is a MS star with a somewhat larger mass as a result of the mass transfer process.  
The orbit significantly shrinks to liberate the orbital gravitational energy that removes the common envelope (equation \ref{eq:EbindEenv}).  

Later, the primary He sub-giant explodes as a stripped-envelope CCSN and leaves a NS remnant. The NS acquires a natal kick velocity. Both the ejected mass at CCSN explosion and the natal kick velocity change the orbit of the binary system, in some cases to become unbound. About $\simeq 2-6 \times 10^9{\rm yr}$ later the secondary star evolves through the Hertzsprung gap and becomes a RG star unless the NS interrupts its evolution. Any mass transfer from the secondary star to the NS at any evolutionary phase might form a (symbiotic) X-ray binary system. For the present study the crucial evolution is the formation of a second CEE phase.

As the secondary evolves along the RG it fills its Roche lobe and the binary system enters an unstable mass transfer. The secondary RG star engulfs the NS and the system experiences a second CEE phase. As it spirals-in inside the RG stellar envelope the NS manages to ejects the entire common envelope. The immediate remnant of the second CEE phase is a close binary system of a NS and a post-RG star, namely an horizontal branch star, and an expanding nebula. 

As the remnant of the core of the RG star contracts and heats-up it might ionize the nebula if its mass is not too low, i.e., its mass should be $\ga 0.3M_\odot$. This evolution leads to a post-RG PN with a central NS-WD binary system. 
Post RG planetary nebulae are rare compared with post-AGB planetary nebulae (e.g., \citealt{Hillwigetal2017, Jonesetal2020, Jonesetal2022, Jonesetal2023}). These are formed by binary interaction (e.g., \citealt{Halletal2013}). The NS-WD PNe that we find here are very rare as we show below.  

\begin{figure}
\centering
\includegraphics[trim=0.3cm 0.0cm 0.0cm 0.0cm ,clip, scale=0.335]{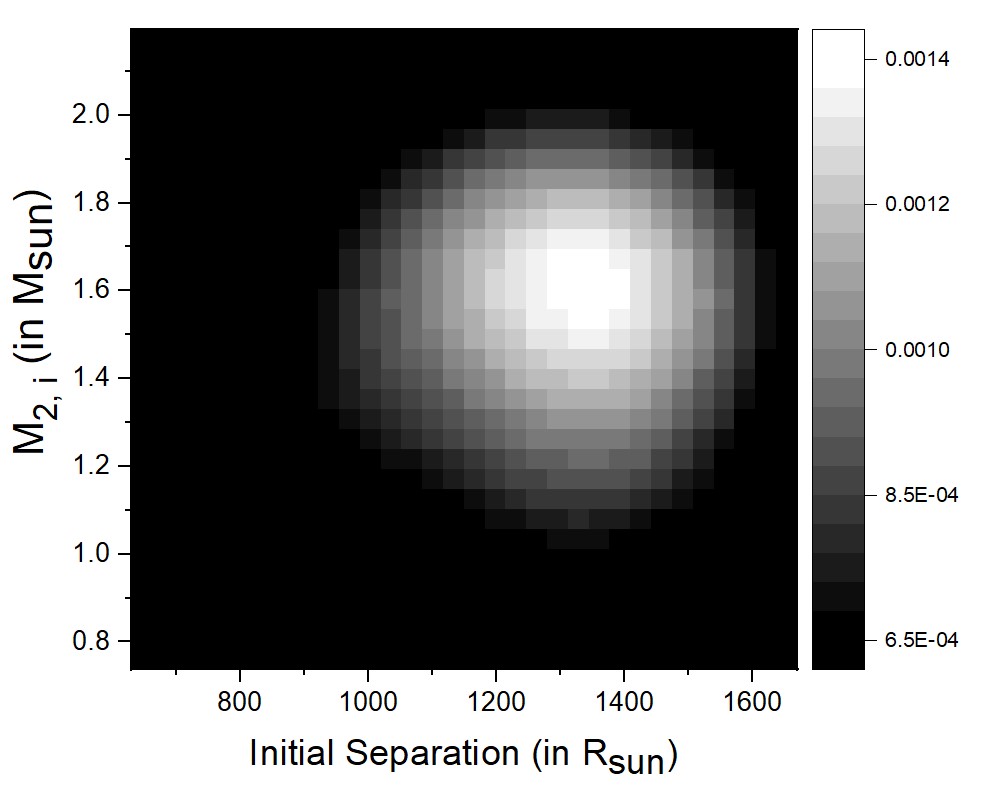}
\caption{The relative number of binary systems that lead to PNe with central NS-WD binary systems in the plane of the initial (ZAMS) secondary mass versus the initial semi-major axis (orbits are eccentric). The gray scale (on the right) represents relative numbers in each rectangle bin in the plot. This plot is made from 52 binary systems. We smoothed their distribution to account for the random nature of the BPS code using a smoothing algorithm with a Gaussian kernel to give the density map in the figure. It represents the data using a continuous probability density curve in two dimensions. In following figures we apply the same smoothing as in this one.     
}
\label{Fig:InitialM2a0}
\end{figure}

We conduct BPS as we describe in section \ref{sec:BinaryEvolution} and find that the initial parameters which lead to the formation of NS-WD PNe under our standard model are as follows. 
(1) The initial primary masses are in the range of $10.0 M_\odot \la M_{1,\rm i} < 12.0M_\odot$. 
(2) The initial secondary masses are in the range of $1 M_\odot \la 
 M_{2,\rm i} \la 2 M_\odot$. We present this range together with the range of the initial semi-major axes in Fig. \ref{Fig:InitialM2a0}. 
(3) The initial semi-major axes are mainly distributed from $\simeq 900 R_\odot$ to $1500 R_\odot$. Binaries with different initial conditions might end up substantially different. They might merge during the first or the second CEE phases, the SN kick may unbind the binary system, or other compact objects may be formed, like two NSs or two WDs.  In appendix \ref{App1} we list the evolutionary phases of three systems that form NS-WD PNe out of the 52 systems that do so in our sample of $10^7$ initial binaries. 

The NS-WD PNe that we study here, i.e., close NS-WD systems with a nebula around them, are very rare (see below). Their formation channel is very delicate and sensitive to parameters that the BPS code includes. Our aim is not to search the parameter space as we do not yet have even one observed candidate for such a system. We only want to show that such systems might form and to present the initial parameters of such systems. Outside this parameter space the system will end differently, i.e., end with the merger of the MS secondary star with the core in the first CEE phase,  the system will end as a wide system without any interaction, or the CCSN explosion unbinds the binary system. In particular we need the first CEE phase to strip the envelope of the primary star such that at CCSN explosion the ejected mass is low. Otherwise the massive ejecta at explosion unbinds the binary system. 

\begin{figure}
\centering
\includegraphics[trim=0.3cm 0.0cm 0.0cm 0.0cm ,clip, scale=0.335]{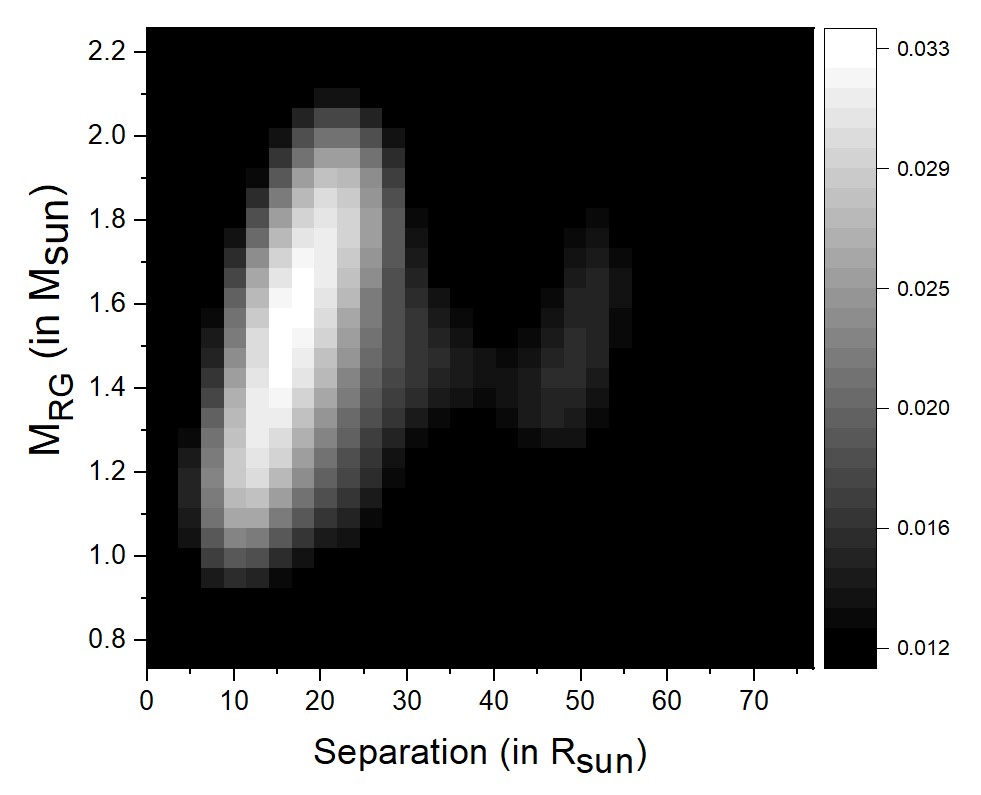}
\caption{The distribution of the secondary stellar mass versus the orbital semi-major axis just before the systems enter the second CEE phase. The secondary star is in the Hertzsprung gap or on the early RG branch. }
\label{Fig:BeforeSecondCEE}
\end{figure}

\begin{figure}
\centering
\includegraphics[trim=0.3cm 0.0cm 0.0cm 0.0cm ,clip, scale=0.335]{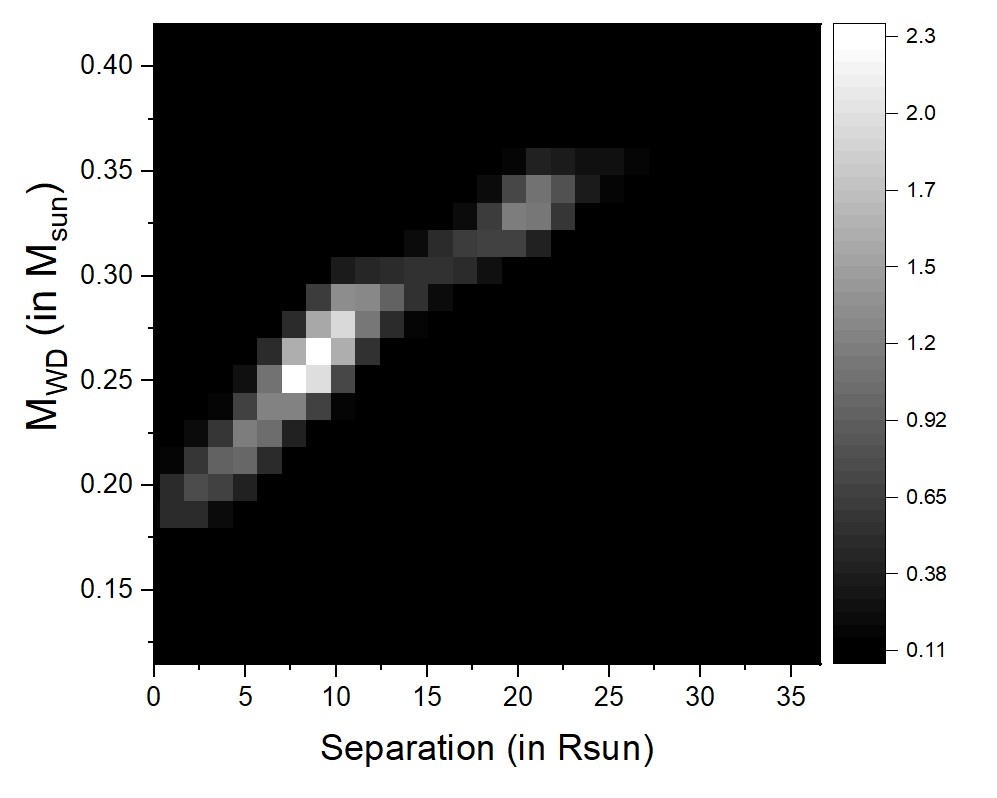}
\caption{Distribution of the final masses of the secondary stars, which are hot horizontal branch stars evolving to become helium WDs, versus their final orbital separations with the NS companions. }
\label{Fig:FinalMassSeparation}
\end{figure}

In Figure \ref{Fig:BeforeSecondCEE} we present the relative number of the binary systems just before the systems experience the second CEE phase in the plane of secondary mass, which is lower than 2 $M_\odot$ and now an RG star, versus the orbital semi-major axis. 
From comparing this secondary mass distribution to the initial one (Fig. \ref{Fig:InitialM2a0}) we learn that the secondary stars accrete negligible amount of mass during the earlier evolution. From the semi-major axes of most systems we learn that the second CEE phase takes place while the secondary system evolves on the Hertzsprung gap or along the early RG star. This implies that the final remnant of the RG star has a small mass. In Figure \ref{Fig:FinalMassSeparation} we present the final masses of He WDs versus the final orbital separations (the final orbits are assumed to be circular, and distributed between 1 and 30 $R_\odot$).   
 
Post-RG stars of mass $M_{\rm WD} \la 0.3 M_\odot$ do not reach high enough effective temperatures to ionize the nebula even if are fully stripped of their envelope and evolve to become a WD (e.g., \citealt{Halletal2013} and as we find here from our BPS simulation). In regular binary system evolution these systems will not form PNe. However, in the NS-WD PN systems that we study here the NS can energies the nebula. The NS accretes mass during the second CEE phase and spins-up and heats up. The NS might be magnetically active and power a pulsar wind nebula inside the expanding nebula. For this, we still term all these systems PNe, even if the final secondary mass is $M_{\rm WD} < 0.3 M_\odot$. We actually have a \textit{PN with a pulsar wind nebula}. 

\cite{Frew2008} finds the PN formation rate in the Milky Way Galaxy to be $\approx 0.4 {\rm yr}^{-1}$, similar to the WD formation rate (for a population study PN formation see, e.g., \citealt{MoeDeMarco2006}). 
From our population study we find that the number of ZAMS stars with mass $1-8M_\odot$ out of all $10^7$ stars are 829529. We assume that each binary system with such a primary mass forms one PN, and each single star in that mass range forms one PN. In calculating the Galactic formation rate of NS-WD PNe and of regular PNe we take the fraction of binary systems to be 0.7 and of single stars to be 0.3. We take a Galactic star formation rate (SFR) of $5 M_\odot \, {\rm yr}^{-1}$ (\citealt{WillemsKolb2004}). Based on our simulation the PN formation rate in Milky Way-like galaxies is $\approx 0.42 {\rm yr}^{-1}$, which is in very good agreement with the value that \cite{Frew2008} infers from observations. 

We also find that the Galactic formation rate of NS-WD PNe is $1.8 \times 10^{-5} {\rm yr}^{-1}$, and the result of our BPS calculations have a statistical error of 14\%. With different prescriptions for uncertain physical processes and parameters, the BPS simulations give different results. For example, if we change the CE parameters to be $\alpha_{\rm CE}=0.1$ and $\lambda=0.1$, the rate of NS-WD PNe changes to zero. We gave the range of 0 - $1.8 \times 10^{-5} {\rm yr}^{-1}$ for the formation rate of He WD-NS PNe. In the 8 different BPS models of \cite{Toonenetal2018}, they provided a range of 0 - $1.3 \times 10^{-5} {\rm yr}^{-1}$ for the Galactic formation rate of NS-He WD systems as gravitational wave (GW) sources (if we use the same SFR and binary fraction) which is in good agreement with our estimated formation rate of close NS-He WD systems (which also can be GW sources) with PNe. However, they did not demonstrate an example of the evolutionary pathway to form NS-He WD systems and did not mention NS-WD PN formation. The different formation rates that \cite{Toonenetal2018} find from different BPS models show that the formation rate of NS-He WD systems is very sensitive to physical parameters in the BPS model, such as the CEE parameters $\alpha_{\rm CE}$ and $\lambda$, and NS kick. Although we find that NS-WD PNe are rare, \cite{Toonenetal2018} and \cite{Breivik2020} present results of simulations with different BPS codes and find rare but non-zero NS-He WD binary population. For other type (CO WD or ONeMg WD) of WD -NS systems, we did not find even one PN from the BPS simulations (no CO or ONeMG WD –NS PNe). In order to form WD- NS PNe, the NS and WD should be formed right after the CE ejection or at least shortly after the giant branch star, but only the evolutionary route shown in this work is the way to for it. Thus, we do not mention other type of systems (CO WD or ONeMg WD - NS systems). \cite{Toonenetal2018} present the evolutionary paths to form CO WD - NS or ONeMg WD - NS systems, but without a PN. However, there may be ONe WD-NS SNRs present directly in the Milky Way and in the Magellanic Clouds (see \cite{bobricketal2022}). Very interestingly, three observed close He WD - NS systems who can merge in a Hubble time are reported in the ATNF Pulsar catalogue  (\citealt{Manchesteretal2005}), and \cite{bobricketal2017} obtained the He WD-NS binary formation rate as $1.0\times 10^{-6} {\rm yr}^{-1}$ which lies in the range of the formation rate given in this work.
We emphasize again the large uncertainties in current stellar and binary evolution that enters BPS codes (see \citealt{AblimitMaeda2018} and \citealt{Belczynskietal2022} for uncertainties in the binary evolution) and that different modeling with different BPS codes might find this rate to be very different.  

Overall the ratio between the number of NS-WD PNe and the number of all PNe is $\approx 1/16000$. Considering that the number of known PNe in the Galaxy is $\simeq 4000$ (e.g., \citealt{Parker2022}; but it will be increasing, e.g., \citealt{Frew2008}), with several hundreds more PNe in the Magellanic clouds (e.g., \citealt{ReidParker2006, ReidParker2010}), at best one of the observed PNe might be a NS-WD PN. The expected number of Galactic PNe according to \cite{Frew2008} is $\simeq 24,000$, and we expect that a few of these might be NS-WD PNe.

\section{Summary}
\label{sec:Summary}

With rapidly growing sky surveys more and more peculiar-rare objects are found. PN surveys include the Hong Kong/AAO/Strasbourg/H$\alpha$ PNe catalogue (HASH, e.g., \citealt{Parkeretal2016, Parker2022, GomezMunozetal2023}) and the INT Photometric H$\alpha$ Survey (IPHAS) of the Northern Galactic Plane (e.g., \citealt{Ritteretal2023}). With enlarged number of observed PNe and with improved studies of individual PNe (e.g., \citealt{DeMarcoetal2022}) we expect that peculiar PNe of different kinds will be found, e.g., a central star with a surviving planet. 
In this study we examined the possibility for very rare PNe with central NS-WD binary systems. This is a very special type of objects, not only for being an interesting peculiar PN, but also because its central NS-WD binary system can be a GW source for detectors like LISA (e.g., \citealt{Tauris2018, Abdusalametal2020}), radio and optical transients (e.g., \citealt{Metzger2012, Zenatietal2019}), and sources of X-ray/gamma-ray bursts (e.g., \citealt{Kingetal2007}).

Fig \ref{Fig:schematic} schematically presents the evolution of a binary system to form a NS-WD PN. We conducted a BPS simulation (we describe the code in section \ref{sec:BinaryEvolution}) to find the properties of such systems, which we present in section \ref{sec:Results}.
Figs. \ref{Fig:InitialM2a0} - \ref{Fig:FinalMassSeparation} present some properties of the binary systems at $t=0$, just before the second CEE phase, and at the end of our evolution, respectively. We simulated $10^7$ MS-MS binary systems and found, for the parameters we use in our BPS code, that the Galactic formation rate of NS-WD PN is $1.8 \times 10^{-5} {\rm yr}^{-1}$ and the PN formation rate in Milky Way-like galaxies is $\approx 0.42 {\rm yr}^{-1}$. 

All the NS-WD PNe we find are post-RG PNe. Many of these have a post-RG star (a hot horizontal branch star) with a mass of $M_\odot \la 0.3 M_\odot$. These post-RG stars to do reach the effective temperatures of $\ga 30,000 {\rm K}$ that are necessary to ionize the nebula. However, we expect the NS to be hot and magnetically active. The NS powers the nebula as well. Therefore, these nebulae will be at least partially ionized (even if not PN by strict definition). We might actually form a PN with a pulsar wind nebula in its inner region.  

Considering the many uncertainties in the binary stellar evolution, like the value of the CEE parameters $\alpha_{\rm CE}$ and $\lambda$, the amount of pre-CEE mass loss, and the natal kick velocities of NSs in close binary systems, the results may have relatively high uncertainty. Different BPS simulations under very different uncertain inputs can yield different fractions of NS-WD PNe (see  Section \ref{sec:Results} for discussion). Nonetheless, our finding is that it is possible that one of the observed PNe in the local group is a NS-WD PN, and that a few NS-WD PNe exist among all PNe in the Galaxy (most that are not detected). Detailed observations of peculiar PNe might reveal strong central X-ray sources similar to magnetically active NSs and to pulsar wind nebulae.

\section*{Acknowledgements}

We thank A. Bobrick, A. Grichener and P. Hashim for their comments on the manuscript.
This work was supported by NSFs. This research was also supported by a grant from the Israel Science Foundation (769/20).

\section*{Data Availability}

The data underlying this article
will be shared on reasonable request to the corresponding author.




\begin{thebibliography}{}\addcontentsline{toc}{section}{References}

\bibitem[\protect\citeauthoryear{Abdusalam et al.}{2020}]{Abdusalametal2020} Abdusalam K., Ablimit I., Hashim P., L{\"u} G.-L., Mardini M.~K., Wang Z.-J., 2020, ApJ, 902, 125. doi:10.3847/1538-4357/abb5a8

\bibitem[\protect\citeauthoryear{Ablimit et al.}{2016}]{Ablimitetal2016} Ablimit I., Maeda K., Li X.-D., 2016, ApJ, 826, 53. doi:10.3847/0004-637X/826/1/53

\bibitem[\protect\citeauthoryear{Ablimit \& Maeda}{2018}]{AblimitMaeda2018} Ablimit I., Maeda K., 2018, ApJ, 866, 151. doi:10.3847/1538-4357/aae378

\bibitem[\protect\citeauthoryear{Ablimit}{2021}]{Ablimit2021} Ablimit I., 2021, PASP, 133, 074201. doi:10.1088/1538-3873/ac025c

\bibitem[\protect\citeauthoryear{Ablimit et al.}{2022}]{Ablimitetal2022} Ablimit I., Podsiadlowski P., Hirai R., Wicker J., 2022, MNRAS, 513, 4802. doi:10.1093/mnras/stac631

\bibitem[\protect\citeauthoryear{Andrews et al.}{2018}]{Andrewsetal2018} Andrews J.~J., Zezas A., Fragos T., 2018, ApJS, 237, 1. doi:10.3847/1538-4365/aaca30

\bibitem[\protect\citeauthoryear{Barker et al.}{2018}]{Bakeretal2018} Barker H., Zijlstra A., De Marco O., Frew D.~J., Drew J.~E., Corradi R.~L.~M., Eisl{\"o}ffel J., et al., 2018, MNRAS, 475, 4504. doi:10.1093/mnras/stx3240

\bibitem[\protect\citeauthoryear{Belczynski et al.}{2008}]{Belczynskietal2008} Belczynski K., Kalogera V., Rasio F.~A., Taam R.~E., Zezas A., Bulik T., Maccarone T.~J., et al., 2008, ApJS, 174, 223. doi:10.1086/521026

\bibitem[\protect\citeauthoryear{Belczynski et al.}{2022}]{Belczynskietal2022} Belczynski K., Romagnolo A., Olejak A., Klencki J., Chattopadhyay D., Stevenson S., Coleman Miller M., et al., 2022, ApJ, 925, 69. doi:10.3847/1538-4357/ac375a

\bibitem[\protect\citeauthoryear{Bode et al.}{1987}]{Bodeetal1987} Bode M.~F., Seaquist E.~R., Frail D.~A., Roberts J.~A., Whittet D.~C.~B., Evans A., Albinson J.~S., 1987, Natur, 329, 519. doi:10.1038/329519a0

\bibitem[\protect\citeauthoryear{Boffin \& Jones}{2019}]{BoffinJones2019} Boffin H.~M.~J., Jones D., 2019, ibfe.book. doi:10.1007/978-3-030-25059-1

\bibitem[\protect\citeauthoryear{Bond}{2000}]{Bond2000} Bond H.~E., 2000, ASPC, 199, 115. doi:10.48550/arXiv.astro-ph/9909516

\bibitem[\protect\citeauthoryear{Bobrick et al.}{2017}]{bobricketal2017} Bobrick A., Davies M.~B., Church R.~P., 2017, MNRAS, 467, 3556. doi:10.1093/mnras/stx312

\bibitem[\protect\citeauthoryear{Bobrick et al.}{2022}]{bobricketal2022} Bobrick A., Zenati Y., Perets H.~B., Davies M.~B., Church R., 2022, MNRAS, 510, 3758. doi:10.1093/mnras/stab3574


\bibitem[\protect\citeauthoryear{Breivik et al.}{2020}]{Breivik2020} Breivik K., Coughlin S., Zevin M., Rodriguez C.~L., Kremer K., Ye C.~S., Andrews J.~J., et al., 2020, ApJ, 898, 71. doi:10.3847/1538-4357/ab9d85

\bibitem[\protect\citeauthoryear{Broekgaarden et al.}{2021}]{Broekgaardenetal2021} Broekgaarden F.~S., Berger E., Neijssel C.~J., Vigna-G{\'o}mez A., Chattopadhyay D., Stevenson S., Chruslinska M., et al., 2021, MNRAS, 508, 5028. doi:10.1093/mnras/stab2716

\bibitem[\protect\citeauthoryear{Chiotellis et al.}{2020}]{Chiotellisetal2020} Chiotellis A., Boumis P., Spetsieri Z.~T., 2020, Galax, 8, 38. doi:10.3390/galaxies8020038

\bibitem[\protect\citeauthoryear{Chiotellis et al.}{2021}]{Chiotellisetal2021} Chiotellis A., Boumis P., Spetsieri Z.~T., 2021, MNRAS, 502, 176. doi:10.1093/mnras/staa3573

\bibitem[\protect\citeauthoryear{Chornay et al.}{2021}]{Chornay2021} Chornay N., Walton N.~A., Jones D., Boffin H.~M.~J., Rejkuba M., Wesson R., 2021, A\&A, 648, A95. doi:10.1051/0004-6361/202140288

\bibitem[\protect\citeauthoryear{Chornay \& Walton}{2022}]{ChornayWalton2022} Chornay N., Walton N.~A., 2022, RNAAS, 6, 177. doi:10.3847/2515-5172/ac8e6c

\bibitem[\protect\citeauthoryear{Ciardullo}{2016}]{Ciardullo2016} Ciardullo R., 2016, IAUFM, 29B, 15. doi:10.1017/S174392131600435X

\bibitem[\protect\citeauthoryear{Cikota et al.}{2017}]{Cikotaetal2017} Cikota A., Patat F., Cikota S., Spyromilio J., Rau G., 2017, MNRAS, 471, 2111. doi:10.1093/mnras/stx1734


\bibitem[\protect\citeauthoryear{Corradi et al.}{2003}]{Corradietal2003} Corradi R.~L.~M., Sch{\"o}nberner D., Steffen M., Perinotto M., 2003, MNRAS, 340, 417. doi:10.1046/j.1365-8711.2003.06294.x

\bibitem[\protect\citeauthoryear{Cox et al.}{2012}]{Coxetal2012} Cox N.~L.~J., Kerschbaum F., van Marle A.-J., Decin L., Ladjal D., Mayer A., Groenewegen M.~A.~T., et al., 2012, A\&A, 537, A35. doi:10.1051/0004-6361/201117910

\bibitem[\protect\citeauthoryear{Davis et al.}{2018}]{Davisetal2018} Davis B.~D., Ciardullo R., Jacoby G.~H., Feldmeier J.~J., Indahl B.~L., 2018, ApJ, 863, 189. doi:10.3847/1538-4357/aad3c4

\bibitem[\protect\citeauthoryear{De Marco}{2009}]{DeMarco2009} De Marco O., 2009, PASP, 121, 316. doi:10.1086/597765

\bibitem[\protect\citeauthoryear{De Marco et al.}{2022}]{DeMarcoetal2022} De Marco O., Akashi M., Akras S., Alcolea J., Aleman I., Amram P., Balick B., et al., 2022, NatAs, 6, 1421. doi:10.1038/s41550-022-01845-2

\bibitem[\protect\citeauthoryear{De Marco et al.}{2015}]{DeMarcoetal2015} De Marco O., Long J., Jacoby G.~H., Hillwig T., Kronberger M., Howell S.~B., Reindl N., et al., 2015, MNRAS, 448, 3587. doi:10.1093/mnras/stv249

\bibitem[\protect\citeauthoryear{de Mink \& Belczynski}{2015}]{deminkBel2015} de Mink S.~E., Belczynski K., 2015, ApJ, 814, 58. doi:10.1088/0004-637X/814/1/58

\bibitem[\protect\citeauthoryear{Drew et al.}{2005}]{Drewetal2005} Drew J.~E., Greimel R., Irwin M.~J., Aungwerojwit A., Barlow M.~J., Corradi R.~L.~M., Drake J.~J., et al., 2005, MNRAS, 362, 753. doi:10.1111/j.1365-2966.2005.09330.x

\bibitem[\protect\citeauthoryear{Dessart et al.}{2006}]{Dessartetal2006} Dessart L., Burrows A., Ott C.~D., Livne E., Yoon S.-C., Langer N., 2006, ApJ, 644, 1063. doi:10.1086/503626

\bibitem[\protect\citeauthoryear{Eldridge et al.}{2018}]{Eldridgeetal2018} Eldridge J.~J., Xiao L., Stanway E.~R., Rodrigues N., Guo N.-Y., 2018, PASA, 35, e049. doi:10.1017/pasa.2018.47

\bibitem[\protect\citeauthoryear{Frank et al.}{2018}]{Frank2018} Frank A., Chen Z., Reichardt T., De Marco O., Blackman E., Nordhaus J., 2018, Galax, 6, 113. doi:10.3390/galaxies6040113

\bibitem[\protect\citeauthoryear{Fragos et al.}{2023}]{Fragosetal2023} Fragos T., Andrews J.~J., Bavera S.~S., Berry C.~P.~L., Coughlin S., Dotter A., Giri P., et al., 2023, ApJS, 264, 45. doi:10.3847/1538-4365/ac90c1

\bibitem[\protect\citeauthoryear{Ferrario}{2012}]{Ferrario2012} Ferrario L., 2012, MNRAS, 426, 2500. doi:10.1111/j.1365-2966.2012.21836.x

\bibitem[\protect\citeauthoryear{Frew}{2008}]{Frew2008} Frew D.~J., 2008, PhDT

\bibitem[\protect\citeauthoryear{Frew \& Parker}{2010}]{FrewParker2010} Frew D.~J., Parker Q.~A., 2010, PASA, 27, 129. doi:10.1071/AS09040

\bibitem[\protect\citeauthoryear{Fryer et al.}{2012}]{Fryeretal2012} Fryer C.~L., Belczynski K., Wiktorowicz G., Dominik M., Kalogera V., Holz D.~E., 2012, ApJ, 749, 91. doi:10.1088/0004-637X/749/1/91

\bibitem[\protect\citeauthoryear{Gagnier \& Pejcha}{2023}]{GagnierPejcha2023} Gagnier D., Pejcha O., 2023, arXiv, arXiv:2302.00691. doi:10.48550/arXiv.2302.00691

\bibitem[\protect\citeauthoryear{Garc{\'\i}a-Segura et al.}{2018}]{GarciaSegura2018} Garc{\'\i}a-Segura G., Ricker P.~M., Taam R.~E., 2018, ApJ, 860, 19. doi:10.3847/1538-4357/aac08c

\bibitem[\protect\citeauthoryear{Garc{\'\i}a-Segura et al.}{2022}]{GarciaSegura2022} Garc{\'\i}a-Segura G., Taam R.~E., Ricker P.~M., 2022, MNRAS, 517, 3822. doi:10.1093/mnras/stac2824

\bibitem[\protect\citeauthoryear{Giacobbo \& Mapelli}{2018}]{GiacobboMapelli2018} Giacobbo N., Mapelli M., 2018, MNRAS, 480, 2011. doi:10.1093/mnras/sty1999

\bibitem[\protect\citeauthoryear{G{\'o}mez-Mu{\~n}oz et al.}{2023}]{GomezMunozetal2023} G{\'o}mez-Mu{\~n}oz M.~A., Bianchi L., Manchado A., 2023, arXiv, arXiv:2304.01970

\bibitem[\protect\citeauthoryear{Greig}{1971}]{Greig1971} Greig W.~E., 1971, A\&A, 10, 161

\bibitem[\protect\citeauthoryear{Guerrero et al.}{2004}]{Guerreroetal2004} Guerrero M.~A., Jaxon E.~G., Chu Y.-H., 2004, AJ, 128, 1705. doi:10.1086/423913

\bibitem[\protect\citeauthoryear{Guerrero et al.}{2019}]{Guerreroetal2019} Guerrero M.~A., Toal{\'a} J.~A., Chu Y.-H., 2019, ApJ, 884, 134. doi:10.3847/1538-4357/ab4256

\bibitem[\protect\citeauthoryear{Guerrero \& De Marco}{2013}]{GuerreroDe Marco2013} Guerrero M.~A., De Marco O., 2013, A\&A, 553, A126. doi:10.1051/0004-6361/201220623

\bibitem[\protect\citeauthoryear{Hall et al.}{2013}]{Halletal2013} Hall P.~D., Tout C.~A., Izzard R.~G., Keller D., 2013, MNRAS, 435, 2048. doi:10.1093/mnras/stt1422

\bibitem[\protect\citeauthoryear{Hamann et al.}{2003}]{Hamannetal2003} Hamann W.-R., Pe{\~n}a M., Gr{\"a}fener G., Ruiz M.~T., 2003, A\&A, 409, 969. doi:10.1051/0004-6361:20031109

\bibitem[\protect\citeauthoryear{Hamers et al.}{2021}]{Hamersetal2021} Hamers A.~S., Rantala A., Neunteufel P., Preece H., Vynatheya P., 2021, MNRAS, 502, 4479. doi:10.1093/mnras/stab287

\bibitem[\protect\citeauthoryear{Heggie}{1975}]{Heggie1975} Heggie D.~C., 1975, MNRAS, 173, 729. doi:10.1093/mnras/173.3.729



\bibitem[\protect\citeauthoryear{Hillwig et al.}{2017}]{Hillwigetal2017} Hillwig T.~C., Frew D.~J., Reindl N., Rotter H., Webb A., Margheim S., 2017, AJ, 153, 24. doi:10.3847/1538-3881/153/1/24

\bibitem[\protect\citeauthoryear{Hillwig et al.}{2016}]{Hillwigetal2016} Hillwig T.~C., Jones D., De Marco O., Bond H.~E., Margheim S., Frew D., 2016, ApJ, 832, 125. doi:10.3847/0004-637X/832/2/125

\bibitem[\protect\citeauthoryear{Hobbs et al.}{2005}]{Hobbsetal2005} Hobbs G., Lorimer D.~R., Lyne A.~G., Kramer M., 2005, MNRAS, 360, 974. doi:10.1111/j.1365-2966.2005.09087.x

\bibitem[\protect\citeauthoryear{Hurley et al.}{2000}]{Hurleyetal2000} Hurley J.~R., Pols O.~R., Tout C.~A., 2000, MNRAS, 315, 543. doi:10.1046/j.1365-8711.2000.03426.x

\bibitem[\protect\citeauthoryear{Hurley et al.}{2002}]{Hurleyetal2002} Hurley J.~R., Tout C.~A., Pols O.~R., 2002, MNRAS, 329, 897. doi:10.1046/j.1365-8711.2002.05038.x

\bibitem[\protect\citeauthoryear{Icke et al.}{1989}]{Icke1989} Icke V., Preston H.~L., Balick B., 1989, AJ, 97, 462. doi:10.1086/114995

\bibitem[\protect\citeauthoryear{Izzard et al.}{2012}]{Izzardetal2012} Izzard R.~G., Hall P.~D., Tauris T.~M., Tout C.~A., 2012, IAUS, 283, 95. doi:10.1017/S1743921312010769

\bibitem[\protect\citeauthoryear{Jacoby et al.}{2021}]{Jacobyetal2021} Jacoby G.~H., Hillwig T.~C., Jones D., Martin K., De Marco O., Kronberger M., Hurowitz J.~L., et al., 2021, MNRAS, 506, 5223. doi:10.1093/mnras/stab2045

\bibitem[\protect\citeauthoryear{Jones et al.}{2019}]{Jonesetal2019} Jones D., Boffin H.~M.~J., Sowicka P., Miszalski B., Rodr{\'\i}guez-Gil P., Santander-Garc{\'\i}a M., Corradi R.~L.~M., 2019, MNRAS, 482, L75. doi:10.1093/mnrasl/sly142

\bibitem[\protect\citeauthoryear{Jones \& Boffin}{2017}]{JonesBoffin2017} Jones D., Boffin H.~M.~J., 2017, NatAs, 1, 0117. doi:10.1038/s41550-017-0117

\bibitem[\protect\citeauthoryear{Jones et al.}{2020}]{Jonesetal2020} Jones D., Boffin H.~M.~J., Hibbert J., Steinmetz T., Wesson R., Hillwig T.~C., Sowicka P., et al., 2020, A\&A, 642, A108. doi:10.1051/0004-6361/202038778

\bibitem[\protect\citeauthoryear{Jones et al.}{2023}] {Jonesetal2023} Jones D., Hillwig T.~C., Reindl, N., arXiv:2304.06355

\bibitem[\protect\citeauthoryear{Jones et al.}{2022}]{Jonesetal2022} Jones D., Munday J., Corradi R.~L.~M., Rodr{\'\i}guez-Gil P., Boffin H.~M.~J., Zak J., Sowicka P., et al., 2022, MNRAS, 510, 3102. doi:10.1093/mnras/stab3736

\bibitem[\protect\citeauthoryear{Kahabka et al.}{2008}]{Kahabkaetal2008} Kahabka P., Haberl F., Pakull M., Millar W.~C., White G.~L., Filipovi{\'c} M.~D., Payne J.~L., 2008, A\&A, 482, 237. doi:10.1051/0004-6361:20078535

\bibitem[\protect\citeauthoryear{Kiel \& Hurley}{2006}]{KielHurley2006} Kiel P.~D., Hurley J.~R., 2006, MNRAS, 369, 1152. doi:10.1111/j.1365-2966.2006.10400.x

\bibitem[\protect\citeauthoryear{King et al.}{2007}]{Kingetal2007} King A., Olsson E., Davies M.~B., 2007, MNRAS, 374, L34. doi:10.1111/j.1745-3933.2006.00259.x

\bibitem[\protect\citeauthoryear{Kruckow et al.}{2016}]{Kruckowetal2016} Kruckow M.~U., Tauris T.~M., Langer N., Sz{\'e}csi D., Marchant P., Podsiadlowski P., 2016, A\&A, 596, A58. doi:10.1051/0004-6361/201629420

\bibitem[\protect\citeauthoryear{Kruckow et al.}{2021}]{Kruckowetal2021} Kruckow M.~U., Neunteufel P.~G., Di Stefano R., Gao Y., Kobayashi C., 2021, ApJ, 920, 86. doi:10.3847/1538-4357/ac13ac

\bibitem[\protect\citeauthoryear{Kronberger et al.}{2014}]{Kronbergeretal2014} Kronberger M., Jacoby G.~H., Acker A., Alves F., Frew D.~J., Goldman D., Guillem P., et al., 2014, apn6.conf, 48

\bibitem[\protect\citeauthoryear{Korol et l.}{2022}]{Koroletal2022} Korol V., Belokurov V., Toonen S., 2022, MNRAS, 515, 1228. doi:10.1093/mnras/stac1686

\bibitem[\protect\citeauthoryear{Kroupa et al.}{1993}]{Kroupaetal1993} Kroupa P., Tout C.~A., Gilmore G., 1993, MNRAS, 262, 545. doi:10.1093/mnras/262.3.545

\bibitem[\protect\citeauthoryear{Kroupa}{2001}]{Kroupa2001} Kroupa P., 2001, MNRAS, 322, 231. doi:10.1046/j.1365-8711.2001.04022.x

\bibitem[\protect\citeauthoryear{Kwitter \& Henry}{2022}]{KwitterHenry2022} Kwitter K.~B., Henry R.~B.~C., 2022, PASP, 134, 022001. doi:10.1088/1538-3873/ac32b1

\bibitem[\protect\citeauthoryear{Kwok}{1983}]{Kwok1983} Kwok S., 1983, IAUS, 103, 293

\bibitem[\protect\citeauthoryear{Manchester et al.}{2005}]{Manchesteretal2005} Manchester R.~N., Hobbs G.~B., Teoh A., Hobbs M., 2005, AJ, 129, 1993. doi:10.1086/428488

\bibitem[\protect\citeauthoryear{Moe \& Di Stefano}{2017}]{MoeDiStefano2017} Moe M., Di Stefano R., 2017, ApJS, 230, 15. doi:10.3847/1538-4365/aa6fb6

\bibitem[\protect\citeauthoryear{Manchado et al.}{2000}]{Manchadoetal2000} Manchado A., Villaver E., Stanghellini L., Guerrero M.~A., 2000, ASPC, 199, 17. doi:10.48550/arXiv.astro-ph/0002073

\bibitem[\protect\citeauthoryear{Mandel et al.}{2021}]{Mandeletal2021} Mandel I., M{\"u}ller B., Riley J., de Mink S.~E., Vigna-G{\'o}mez A., Chattopadhyay D., 2021, MNRAS, 500, 1380. doi:10.1093/mnras/staa3390

\bibitem[\protect\citeauthoryear{Mandel \& Broekgaarden}{2022}]{MandelBroekgaarden2022} Mandel I., Broekgaarden F.~S., 2022, LRR, 25, 1. doi:10.1007/s41114-021-00034-3

\bibitem[\protect\citeauthoryear{Mapelli et al.}{2019}]{Mapelli2019} Mapelli M., Giacobbo N., Santoliquido F., Artale M.~C., 2019, MNRAS, 487, 2. doi:10.1093/mnras/stz1150

\bibitem[\protect\citeauthoryear{Maitra \& Haberl}{2022}]{MaitraHaberl2022} Maitra C., Haberl F., 2022, A\&A, 657, A26. doi:10.1051/0004-6361/202142159

\bibitem[\protect\citeauthoryear{Marchant et al.}{2021}]{Marchantetal2021} Marchant P., Pappas K.~M.~W., Gallegos-Garcia M., Berry C.~P.~L., Taam R.~E., Kalogera V., Podsiadlowski P., 2021, A\&A, 650, A107. doi:10.1051/0004-6361/202039992

\bibitem[\protect\citeauthoryear{Metzger}{2012}]{Metzger2012} Metzger B.~D., 2012, MNRAS, 419, 827. doi:10.1111/j.1365-2966.2011.19747.x

\bibitem[\protect\citeauthoryear{Miszalski et al.}{2008}]{Miszalskietal2008} Miszalski B., Parker Q.~A., Acker A., Birkby J.~L., Frew D.~J., Kovacevic A., 2008, MNRAS, 384, 525. doi:10.1111/j.1365-2966.2007.12727.x

\bibitem[\protect\citeauthoryear{Miszalski et al.}{2009}]{Miszalskietal2009} Miszalski B., Acker A., Parker Q.~A., Moffat A.~F.~J., 2009, A\&A, 505, 249. doi:10.1051/0004-6361/200912176

\bibitem[\protect\citeauthoryear{Moe \& De Marco}{2006}]{MoeDeMarco2006} Moe M., De Marco O., 2006, ApJ, 650, 916. doi:10.1086/506900

\bibitem[\protect\citeauthoryear{Morris}{1981}]{Morris1981} Morris M., 1981, ApJ, 249, 572. doi:10.1086/159317

\bibitem[\protect\citeauthoryear{Morris}{1987}]{Morris1987} Morris M., 1987, PASP, 99, 1115. doi:10.1086/132089

\bibitem[\protect\citeauthoryear{Munari et al.}{2013}]{Munarietal2013} Munari U., Corradi R.~L.~M., Siviero A., Baldinelli L., Maitan A., 2013, A\&A, 558, A2. doi:10.1051/0004-6361/201321883

\bibitem[\protect\citeauthoryear{Oh et al.}{2023}]{Ohetal2023} Oh M., Fishbach M., Kimball C., Kalogera V., Ye C., 2023, arXiv, arXiv:2303.06081. doi:10.48550/arXiv.2303.06081

\bibitem[\protect\citeauthoryear{Ondratschek et al.}{2022}]{Ondratschek2022} Ondratschek P.~A., R{\"o}pke F.~K., Schneider F.~R.~N., Fendt C., Sand C., Ohlmann S.~T., Pakmor R., et al., 2022, A\&A, 660, L8. doi:10.1051/0004-6361/202142478

\bibitem[\protect\citeauthoryear{Olejak et al.}{2021}]{Olejaketal2021} Olejak A., Belczynski K., Ivanova N., 2021, A\&A, 651, A100. doi:10.1051/0004-6361/202140520

\bibitem[\protect\citeauthoryear{{\"O}pik}{1924}]{opik1924} {\"O}pik E., 1924, PTarO, 25, 1

\bibitem[\protect\citeauthoryear{Paczy{\'n}ski}{1971}]{Paczynski1971} Paczy{\'n}ski B., 1971, ARA\&A, 9, 183. doi:10.1146/annurev.aa.09.090171.001151

\bibitem[\protect\citeauthoryear{Parker}{2022}]{Parker2022} Parker Q.~A., 2022, FrASS, 9, 895287. doi:10.3389/fspas.2022.895287

\bibitem[\protect\citeauthoryear{Parker et al.}{2016}]{Parkeretal2016} Parker Q.~A., Boji{\v{c}}i{\'c} I.~S., Frew D.~J., 2016, JPhCS, 728, 032008. doi:10.1088/1742-6596/728/3/032008

\bibitem[\protect\citeauthoryear{Parker et al.}{2005}]{Parkeretal2005} Parker Q.~A., Phillipps S., Pierce M.~J., Hartley M., Hambly N.~C., Read M.~A., MacGillivray H.~T., et al., 2005, MNRAS, 362, 689. doi:10.1111/j.1365-2966.2005.09350.x

\bibitem[\protect\citeauthoryear{Pejcha et al.}{2012}]{Pejchaetal2012} Pejcha O., Thompson T.~A., Kochanek C.~S., 2012, MNRAS, 424, 1570. doi:10.1111/j.1365-2966.2012.21369.x

\bibitem[\protect\citeauthoryear{Podsiadlowski et al.}{2004}]{Podsiadlowskietal2004} Podsiadlowski P., Langer N., Poelarends A.~J.~T., Rappaport S., Heger A., Pfahl E., 2004, ApJ, 612, 1044. doi:10.1086/421713

\bibitem[\protect\citeauthoryear{Portegies et al.}{1998}]{PortegiesYun1998} Portegies Zwart S.~F., Yungelson L.~R., 1998, A\&A, 332, 173. doi:10.48550/arXiv.astro-ph/9710347

\bibitem[\protect\citeauthoryear{Reid \& Parker}{2006}]{ReidParker2006} Reid W.~A., Parker Q.~A., 2006, MNRAS, 365, 401. doi:10.1111/j.1365-2966.2005.09757.x

\bibitem[\protect\citeauthoryear{Reid \& Parker}{2010}]{ReidParker2010} Reid W.~A., Parker Q.~A., 2010, MNRAS, 405, 1349. doi:10.1111/j.1365-2966.2010.16635.x

\bibitem[\protect\citeauthoryear{Riley et al.}{2022}]{Rileyetal2022} Riley J., Agrawal P., Barrett J.~W., Boyett K.~N.~K., Broekgaarden F.~S., Chattopadhyay D., Gaebel S.~M., et al., 2022, ApJS, 258, 34. doi:10.3847/1538-4365/ac416c

\bibitem[\protect\citeauthoryear{R{\"o}pke \& De Marco}{2023}]{ropkede2023} R{\"o}pke F.~K., De Marco O., 2023, LRCA, 9, 2. doi:10.1007/s41115-023-00017-x

\bibitem[\protect\citeauthoryear{Ritter et al.}{2023}]{Ritteretal2023} Ritter A., Parker Q.~A., Sabin L., Le D{\^u} P., Mulato L., Patchick D., 2023, MNRAS, 520, 773. doi:10.1093/mnras/stac2896


\bibitem[\protect\citeauthoryear{Sabin et al.}{2014}]{Sabinetal2014} Sabin L., Parker Q.~A., Corradi R.~L.~M., Guzman-Ramirez L., Morris R.~A.~H., Zijlstra A.~A., Boji{\v{c}}i{\'c} I.~S., et al., 2014, MNRAS, 443, 3388. doi:10.1093/mnras/stu1404


\bibitem[\protect\citeauthoryear{Sahai et al.}{2011}]{Sahai2011} Sahai R., Morris M.~R., Villar G.~G., 2011, AJ, 141, 134. doi:10.1088/0004-6256/141/4/134

\bibitem[\protect\citeauthoryear{Sana et al.}{2012}]{Sanaetal2012} Sana H., de Mink S.~E., de Koter A., Langer N., Evans C.~J., Gieles M., Gosset E., et al., 2012, Sci, 337, 444. doi:10.1126/science.1223344

\bibitem[\protect\citeauthoryear{Sch{\"o}nberner et al.}{2007}]{Schonberneretal2007} Sch{\"o}nberner D., Jacob R., Steffen M., Sandin C., 2007, A\&A, 473, 467. doi:10.1051/0004-6361:20077437

\bibitem[\protect\citeauthoryear{Stanghellini et al.}{2002}]{Stanghellinietal2002} Stanghellini L., Villaver E., Manchado A., Guerrero M.~A., 2002, ApJ, 576, 285. doi:10.1086/341340

\bibitem[\protect\citeauthoryear{Soker}{1997}]{Soker1997} Soker N., 1997, ApJS, 112, 487. doi:10.1086/313040

\bibitem[\protect\citeauthoryear{Soker}{2006}]{Soker2006} Soker N., 2006, PASP, 118, 260. doi:10.1086/498829

\bibitem[\protect\citeauthoryear{Soker \& Rappaport}{2000}]{SokerRappaport2000} Soker N., Rappaport S., 2000, ApJ, 538, 241. doi:10.1086/309112

\bibitem[\protect\citeauthoryear{Soker \& Rappaport}{2001}]{SokerRappaport2001} Soker N., Rappaport S., 2001, ApJ, 557, 256. doi:10.1086/321669

\bibitem[\protect\citeauthoryear{Souropanis et al.}{2023}]{Souropanis2023} Souropanis D., Chiotellis A., Boumis P., Jones D., Akras S., 2023, MNRAS, 521, 1808. doi:10.1093/mnras/stad564

\bibitem[\protect\citeauthoryear{Tauris}{2018}]{Tauris2018} Tauris T.~M., 2018, PhRvL, 121, 131105. doi:10.1103/PhysRevLett.121.131105

\bibitem[\protect\citeauthoryear{Tanikawa et al.}{2022}]{Tanikawaetal2022} Tanikawa A., Yoshida T., Kinugawa T., Trani A.~A., Hosokawa T., Susa H., Omukai K., 2022, ApJ, 926, 83. doi:10.3847/1538-4357/ac4247

\bibitem[\protect\citeauthoryear{Toonen et al.}{2012}]{Toonenetal2012} Toonen S., Nelemans G., Portegies Zwart S., 2012, A\&A, 546, A70. doi:10.1051/0004-6361/201218966

\bibitem[\protect\citeauthoryear{Toonen et al.}{2018}]{Toonenetal2018} Toonen S., Perets H.~B., Igoshev A.~P., Michaely E., Zenati Y., 2018, A\&A, 619, A53. doi:10.1051/0004-6361/201833164

\bibitem[\protect\citeauthoryear{Trani et al.}{2022}]{Tranietal2022} Trani A.~A., Rastello S., Di Carlo U.~N., Santoliquido F., Tanikawa A., Mapelli M., 2022, MNRAS, 511, 1362. doi:10.1093/mnras/stac122

\bibitem[\protect\citeauthoryear{Tsebrenko \& Soker}{2013}]{TsebrenkoSkoer2013} Tsebrenko D., Soker N., 2013, MNRAS, 435, 320. doi:10.1093/mnras/stt1301

\bibitem[\protect\citeauthoryear{Tsebrenko \& Soker}{2015}]{TsebrenkoSkoer2015} Tsebrenko D., Soker N., 2015, MNRAS, 447, 2568. doi:10.1093/mnras/stu2567

\bibitem[\protect\citeauthoryear{Tweedy \& Kwitter}{1994}]{TweedyKwitter1994} Tweedy R.~W., Kwitter K.~B., 1994, AJ, 108, 188. doi:10.1086/117057

\bibitem[\protect\citeauthoryear{van Son et al.}{2022}]{vansonetal2022} van Son L.~A.~C., de Mink S.~E., Callister T., Justham S., Renzo M., Wagg T., Broekgaarden F.~S., et al., 2022, ApJ, 931, 17. doi:10.3847/1538-4357/ac64a3

\bibitem[\protect\citeauthoryear{van Zeist et al.}{2023}]{vanetal2023} van Zeist W.~G.~J., Eldridge J.~J., Tang P.~N., 2023, arXiv, arXiv:2301.06888. doi:10.48550/arXiv.2301.06888

\bibitem[\protect\citeauthoryear{Vink et al.}{2001}]{Vinketal2001} Vink J.~S., de Koter A., Lamers H.~J.~G.~L.~M., 2001, A\&A, 369, 574. doi:10.1051/0004-6361:20010127

\bibitem[\protect\citeauthoryear{Vink \& de Koter}{2005}]{VinkdeKoter2005} Vink J.~S., de Koter A., 2005, A\&A, 442, 587. doi:10.1051/0004-6361:20052862

\bibitem[\protect\citeauthoryear{Wang et al.}{2015}]{wangetal2015} Wang B., Ma X., Liu D.-D., Liu Z.-W., Wu C.-Y., Zhang J.-J., Han Z., 2015, A\&A, 576, A86. doi:10.1051/0004-6361/201425294

\bibitem[\protect\citeauthoryear{Webbink}{1984}]{Webbink1984} Webbink R.~F., 1984, ApJ, 277, 355. doi:10.1086/161701

\bibitem[\protect\citeauthoryear{Wesson et al.}{2008}]{Wessonetal2008} Wesson R., Barlow M.~J., Corradi R.~L.~M., Drew J.~E., Groot P.~J., Knigge C., Steeghs D., et al., 2008, ApJL, 688, L21. doi:10.1086/594366

\bibitem[\protect\citeauthoryear{Willems \& Kolb}{2004}]{WillemsKolb2004} Willems B., Kolb U., 2004, A\&A, 419, 1057. doi:10.1051/0004-6361:20040085

\bibitem[\protect\citeauthoryear{Zenati et al.}{2019}]{Zenatietal2019} Zenati Y., Perets H.~B., Toonen S., 2019, MNRAS, 486, 1805. doi:10.1093/mnras/stz316

\end{thebibliography}


 




\appendix

\setcounter{figure}{0}                  
\setcounter{table}{0}

\section{Examples of evolutionary phases}
\label{App1}

In Tables \ref{Table1} - \ref{Table3} we list the evolutionary phases of three different systems that form NS-WD PNe. These are three examples out of the 52 systems that we find to form NS-WD PNe when we start with $10^7$ binaries on the ZAMS in one BPS model. 
These examples can serve for comparison with future similar BPS simulations with different (or same) inputs or/and future observations. 
Other example are available from the authors upon request. 
\begin{table}
\begin{center}
\caption{Evolutionary phases to form a NS-WD PN: Example 1. }
\label{Table1}
\begin{tabular}{|cccccc|}
\hline
Time& $M_1$   &k1 & k2 &$M_2$&$a$ \\
Myr &$M_\odot$&       &  &$M_\odot$&$R_\odot$ \\ 
\hline
 0.00 &11.30 & MS & MS & 1.20 & 1318.83     \\
 19.81&11.05 & CHeB & MS & 1.20 &1335.85  \\
 22.12&10.82 & CHeB(RLOF)&MS&1.201&876.38   \\
 22.12&3.32& CEE&CEE&1.201&26.37           \\
 22.12&3.32& HeHG(CCSN) &MS&1.201&26.37    \\
22.23&1.47 & NS & MS & 1.201 &38.78       \\
 5591.94&1.47 & NS & HG & 1.201 &37.37      \\
 5886.45&1.47 & NS & RG &1.201&31.71        \\
6296.16&1.471& NS &RG(RLOF)&1.197&29.15     \\
 6296.16&1.471& CEE & CEE &0.24& 2.18        \\
 6296.16 &1.471 & NS & WD& 0.24& 2.18        \\
\hline
\end{tabular}
\end{center}
Note: The first column is the evolution time starting at ZAMS. The second and fifth columns are the masses of the primary and secondary stars, respectively. The third and fourth columns are the evolutionary phases of the primary and secondary stars, respectively, and some processes the binary system experiences. The last column is the orbital separation. 
Abbreviations: CCSN: core collapse supernova; CEE: common envelope evolution; CHeB: core helium burning; HeHG: Hertzsprung Gap stripped-envelope helium star (helium sub-giant); MS: main sequence (star); NS: neutron star; RG: red giant; RLOF: Roche lobe overflow; WD: white dwarf.
\end{table}
\begin{table}
\begin{center}
\caption{Evolutionary phases to form a NS-WD PN: Example 2. }
\label{Table2}
\begin{tabular}{|cccccc|}
\hline
Time & $M_1$ & k1 & k2 & $M_2$ & $a$ \\
Myr &$M_\odot$&       &  &$M_\odot$&$R_\odot$ \\
\hline
  0.00 &10.90 & MS & MS & 1.66 & 950.30       \\
 21.11&10.66 & CHeB & MS & 1.66 &818.05     \\
 23.55&10.47 & CHeB(RLOF)&MS&1.661&781.55   \\
 23.55&3.17& CEE&CEE&1.661&29.76           \\
 23.55&3.17& HeHG(CCSN) &MS&1.661&29.76     \\
23.70&1.45 & NS & MS & 1.663 &45.43          \\
2000.37&1.45 & NS & HG & 1.663 &45.41         \\
2033.12&1.45& NS & RG &1.663&40.65          \\
 2103.57&1.45& NS &RG(RLOF)&1.661&33.03     \\
 2103.57&1.45& CEE & CEE &0.31& 1.58       \\
 2475.78 &1.45 & NS & WD& 0.31& 1.58     \\
\hline
\end{tabular}
\end{center}
Note: Similar to table\ref{Table1} but for a different binary system.  
\end{table}
\begin{table}
\begin{center}
\caption{Evolutionary phases to form a NS-WD PN: Example 3. }
\label{Table3}
\begin{tabular}{|cccccc|}
\hline
Time & $M_1$ & k1 & k2 & $M_2$ & $a$ \\
Myr &$M_\odot$&       &  &$M_\odot$&$R_\odot$ \\
\hline
0.00 &10.60 & MS & MS & 1.58&1182.58     \\
24.21&10.18 & CHeB & MS & 1.58 &1187.04   \\
24.81&10.17 & CHeB(RLOF)&MS&1.58&825.98    \\
24.81&3.05& CEE&CEE&1.58&32.36            \\
24.81&3.05& HeHG(CCSN) &MS&1.58&32.36      \\
24.93&1.4 & NS & MS & 1.58 &107.72        \\
2334.38&1.4 & NS & HG & 1.58 &107.76        \\
2370.94&1.4& NS & RG &1.58&105.51           \\
2475.78&1.4& NS &RG(RLOF)&1.579&46.24       \\
2475.78&1.4& CEE & CEE &0.30& 2.59         \\
2475.78 &1.4 & NS & WD& 0.30& 2.59         \\

\hline
\end{tabular}
\end{center}
Note: Similar to table\ref{Table1} but for a different binary system.  
\end{table}


\bsp	
\label{lastpage}
\end{document}